\documentclass[aps,prb,twocolumn,superscriptaddress,floatfix,longbibliography,citeautoscript, 10pt]{revtex4-2}

\usepackage{amsmath,amssymb} 
\usepackage{bm} 
\usepackage{graphicx} 
\usepackage{comment} 
\usepackage{mathtools}
\usepackage{orcidlink}

\usepackage{textcomp} 

\usepackage{enumitem}
\setlist{noitemsep,leftmargin=*,topsep=0pt,parsep=0pt}

\usepackage{xcolor} 
\definecolor{lightgray}{gray}{0.6}
\definecolor{medgray}{gray}{0.4}

\usepackage{hyperref}
\hypersetup{
colorlinks=true,
urlcolor= blue,
citecolor=blue,
linkcolor= blue,
}

\usepackage{aas_macros}

\newif\ifptitle
\newif\ifpnumber
\newcounter{para}
\newcommand\ptitle[1]{\par\refstepcounter{para}
{\ifpnumber{\noindent\textcolor{lightgray}{\textbf{\thepara}}\indent}\fi}
{\ifptitle{\textbf{\color{blue}[{#1}]}}\fi}}

\newcommand{\rrkm}{\tau_{\text{RRKM}}}
\newcommand{\response}[1]{#1}
\newcommand{\final}[1]{#1}
\newcommand{\mytitle}{Limits of Statistical Models of Ultracold Complex Lifetimes}

\usepackage{etoolbox}
\AtBeginEnvironment{thebibliography}{\interlinepenalty=10000}

\begin{document}

\title{\mytitle}

\author{Kevin B. Xu\orcidlink{0000-0002-6929-2852}}
\affiliation{JILA and Department of Physics, University of Colorado, Boulder, Colorado 80309, USA}

\author{John L. Bohn\orcidlink{0000-0002-4657-0880}}
\affiliation{JILA, NIST, and Department of Physics, University of Colorado, Boulder, Colorado 80309, USA}

\date{\today}

\begin{abstract}
The puzzle of ``sticky collisions,'' in which molecular collision complexes exhibit unexpectedly long lifetimes, remains an unresolved mystery. A central challenge to solving this mystery is that traditional close-coupling calculations remain limited by the vast computational cost needed to take into account all the degrees of freedom involved in the collision. In this work, we propose a statistical model designed to simulate the result of full close-coupling calculations, with the goal of collecting statistics about reasonable lifetimes of collision complexes. To do so, we numerically sample resonances using random matrix theory and utilize results from quantum defect theory to calculate scattering properties and lifetimes. We find that in the limit of dense resonances, our theory agrees well with the Rice-Ramsperger-Kassel-Markus (RRKM) prediction, whereas in the limit of sparse resonances, the physics is governed by threshold behavior rather than resonant effects. By comparing these predictions to experimental results in two limits, we \response{clarify the link between density of states and the validity of the lifetimes they predict.} 


\end{abstract}

\maketitle
\section{\label{sec:intro}Introduction}
\ptitle{introduce the issue of sticky collisions}
The platform of ultracold molecules promises many powerful tools to probe novel physics, and significant efforts in recent years have focused on creating colder and more coherent molecular gases \cite{softely_review, cornish2024quantum, Langen_2024, karman2024ultracold, heller_review}. However, these efforts have faced a persistent challenge: large two-body losses that limit both cooling techniques and lifetimes of experiments \cite{ospelkaus_2012}. This fast rate of two-body loss is now believed to be a result of molecular collisions forming long-lived complexes which are subsequently lost from the trap through mechanisms such as photoassociation \cite{photoassociation, Gersema_2021, Liu_nature}. Preventing the molecules from reaching short range has thus become a prevailing theme in the field, and recent advancements in suppressing short-range collisions through the use of microwave shielding saw the creation of the first molecular Bose-Einstein condensate, a historic landmark for the field \cite{Bigagli_2024}. However, despite the experimental progress, much of the theory behind collision complexes is still unknown. The mechanism behind the long lifetimes of the complexes remains particularly mysterious, with large discrepancies between theoretical predictions and experimental observations. Specifically, several experiments \cite{nichols_PhysRevX.12.011049, Gersema_2021, bause_prr} have measured complex lifetimes which are orders of magnitude longer than expected by theory estimates.

Part of the challenge is that theoretical investigations of these lifetimes are extremely difficult to evaluate directly in the time domain with quantum mechanics, using wavepacket propagation under the Schrödinger equation. This is because extremely small time steps would be required to track the multidimensional wavepacket in the depths of a deep potential energy surface, but the total time of propagation must remain extremely long to capture the entire collision. Ultimately, even with sufficient computational resources, it is not feasible to converge such calculations. As a simpler alternative, classical trajectory simulations have been carried out in several cases \cite{heller_classical, chaos_croft, heller_review, lide_thesis, man_classical, klos_classical}, yet still without accounting quantitatively for the empirically long lifetimes measured by some experiments.

As an alternative to directly computing time evolution, quantum scattering calculations in the energy domain show promise in estimating lifetimes.  The technology for performing these calculations is well-established, and is often limited only by computing power. For example, estimates of the distribution of lifetimes in $\rm{K}+\rm{KRb}$ cold collisions have recently been obtained \cite{Croft2017_PES, kendrick_KKRb}, considering the ro-vibrational degrees of freedom of the $\rm{K_2Rb}$ complex while neglecting spin degrees of freedom.  
\response{Conversely, a theoretical study of spin-changing $\rm{Rb}+\rm{KRb}$ collisions included all spin degrees of freedom while treating KRb as a rigid rotor and neglecting vibrational motion \cite{timur_K2KRb}.}
However, assembling all relevant degrees of freedom in a single calculation remains an unattainable goal in the present, given the current amount of computing power.

\response{Against this backdrop, we present the following approach. While performing a large-scale close-coupling calculation is prohibitively difficult, we can nevertheless contemplate a {\it simulation} of such a calculation.} The model in this simulation assumes that the quantities (resonances) characterizing the close-coupling calculation follow certain statistical distributions defined by a few parameters. The true quantities from the actual calculation represent just one point in this statistical distribution, but randomly drawing from these distributions allows us to simulate other hypothetical systems governed by the same statistical parameters. While this may not produce the results for any real system, this importantly allows us to {\it constrain} the possible results of close-coupling calculations based on statistics, at least within the model's assumptions. \response{Ultimately, our results provide a test of whether a full close-coupling calculation, with all degrees of freedom, is able to reproduce the long lifetimes observed in experiments.}



More specifically, the simulation is based on a random matrix approach.  Drawing on a concept in \cite{mayle_PhysRevA.85.062712}, we generate ensembles of resonances at energies characterized by a given density of states, along with given bound-free couplings that govern the mean resonance widths. Each realization of a spectrum then yields an energy-dependent scattering matrix $S(E)$ containing all of the resonance information. This $S$-matrix is the quantity that is thought of as a proxy for the actual, lengthy close-coupling calculation. From the $S$-matrix, a lifetime can be evaluated via the Wigner-Smith time delay. 
By using this model, we can evaluate the complete influence of a dense forest of resonances without needing to obtain the precise energies and widths from the full close-coupling calculation. Ultimately, this will enable us to construct a distribution of possible time delays for a given molecular species. We note that this is similar to the rationale that has been used previously to understand the statistical distribution of scattering lengths in atomic cold collisions \cite{gribakin_pra, wang_bohn}.

In a second step, we convert the information in $S(E)$ into explicitly time-dependent wavepackets.
To this end, we construct single-channel wavepackets at large intermolecular spacing, whose short-range behavior is modeled solely by the energy dependence of the S-matrix. The time delays, as explicitly seen in the reflected wavepackets, are generally consistent with the time delay as computed in the energy domain. \response{This provides a useful consistency check between the energy- and time-domain representations, and it also illustrates how the energy-dependent time delay is manifested in the propagation and spatial structure of the scattered wavepacket.}

As a result of this analysis, we find that time delays cluster around characteristic values, which depend on the relative scale of the density of resonant states, $\rho$, as compared to the temperature $T$ of the gas.  In the limit of dense resonances ($\rho k_BT \gg 1$), many resonances are accessible to the collision partners. Averaging over resonances is therefore reasonable, and the characteristic lifetime is given by the familiar Rice-Ramsperger-Kassel-Markus (RRKM) expression $\rrkm = 2 \pi \hbar \rho$ for a single open channel. This is the conventional value that arises from random matrix theory \cite{mayle_PhysRevA.85.062712, Mitchell_2010, transition_state_theory, rrkm, rice_ramsperger} and is often used as a basis of comparison for other results.

In the opposing limit of sparse resonances  ($\rho k_BT \ll 1$), it is unlikely that the collision partners experience any resonance at all. In this case, a more appropriate characteristic time scale is given by threshold effects. \response{This results in time delays on the order of ${\bar a} \sqrt{2m / k_BT}$, where $m$ is the reduced mass and ${\bar a}$ is the Gribakin-Flambaum mean scattering length, characteristic of the long-range interaction between scattering partners \cite{gribakin_pra}. Some experiments to date have measured species whose complex lifetimes lie in this sparse limit, yet the empirical lifetimes tend to be far larger than our predicted time scale.} The results presented here therefore point to the limits of random-matrix theory ideas in describing complex lifetimes.

\ptitle{outline sections}
Consequently, the central result of this work is that collision complex lifetimes fall into two distinct regimes: a dense regime where statistical theories of resonances apply, and a sparse regime where threshold physics dominates. \response{In the latter case, lifetimes do not appear to be governed primarily by resonances, which presents difficulties for statistical approaches. }

The remainder of this paper is organized as follows. Sec.~\ref{sec:theory} reviews the statistical theory of lifetimes for collision complexes. Sec.~\ref{sec:results} presents the results of our simulations, focusing on the two limits. Sec.~\ref{sec:experiments} then comments on how our results apply to experimentally measured complex lifetimes. Finally, we conclude with a discussion in Sec.~\ref{sec:conclusion}.

\section{\label{sec:theory}Theory}
\ptitle{outline the idea}
\response{The key assumption of our model is that, even if the most accurate close-coupling calculation were performed on a given scattering system, the result of such a calculation would nevertheless consist of a scattering matrix $S(E)$ that is characterized by resonances. The purpose of our simplified model is to plausibly reproduce the form of $S(E)$, including a realistic distribution of resonances. The model is therefore firmly grounded in random matrix theory (RMT), which has been shown to produce resonance distributions consistent with those obtained from close-coupling calculations \cite{Croft_2017}.}

\subsection{Statistical Theory of Resonances}

\response{The present model is based on a previously developed RMT approach to resonant scattering \cite{mayle_PhysRevA.85.062712, Croft23}, which we briefly summarize in this section. To begin with, we consider two-body collisions in an ultracold gas with a single open scattering channel.} The resonant spectrum of the collisions is assumed to follow RMT and is entirely characterized by two parameters: (1) a density of states $\rho$  corresponding to mean level spacing $d = 1/ \rho$, and (2) \final{a mean-square coupling to the continuum $\nu^2$}. Different species of molecules are  uniquely determined by their values of $d$ and $\nu^2$ in our model.

The resonance energies $E_{\mu}$ are selected randomly such that their mean level spacing is $d$. Because of the RMT assumption, the normalized 
nearest-neighbor level splittings $s_\mu=(E_{\mu+1}-E_\mu)/d$ satisfy the Wigner-Dyson distribution given by
\begin{equation}
\label{eq:wigner-dyson}
    P_{w}(s) = \frac{\pi}{2}se^{-\pi s^2 / 4}.
\end{equation}
In the model, each resonance is furthermore coupled to a single open continuum channel via a random variable $W_{\mu}$, which ultimately governs the width of the resonance. The values of $W_\mu$ are selected from a zero-mean Gaussian distribution satisfying
\begin{equation}
\label{eq:couplings}
    \langle W_i W_j \rangle = \delta_{ij} \nu^2.
\end{equation}
More generally, these couplings are conventionally parametrized in units of the mean level spacing \cite{Mitchell_2010}, which defines a dimensionless parameter
\begin{equation}
\label{eq:dimensionlessx}
x = \frac{(\pi\nu)^2}{d}.
\end{equation}
In these units, the limit $x \ll 1$ defines the region of narrow isolated resonances.  In the opposite limit, $x \gg 1$, the spectrum consists of a small number of extremely broad resonances (i.e., wider than the mean level spacing), along with a collection of narrow resonances just as in the $x \ll 1$ limit \cite{Mitchell_2010, Croft23}. \response{At the special point $x=1$, the average resonance width is maximized and the transmission probability between the short- and long-range regions}
\begin{equation}
\label{eq:transmission}
    T = \frac{4x}{(1+x)^2}
\end{equation}
reaches unity \cite{transition_state_theory}. \response{The case $T=1$ furthermore corresponds to the RRKM limit of transition-state theory and implies that all flux reaching short range forms a complex (the universal-loss limit)} \cite{mayle_PhysRevA.85.062712, julienne_background, Mitchell_2010}. For most species under consideration, the transmission probability has been inferred to be of order unity from two-body loss rates measured in experiments \cite{Gregory_2020, nichols_PhysRevX.12.011049}, and for this reason we will mostly assume ${x}=1$ in the applications to follow. \response{Note that because $x$ is held fixed, increasing the mean level spacing $d$ also increases the resonance widths, approximately preserving the degree of resonance overlap.}


\subsection{Resonant Scattering}
The resonances thus defined are assumed to belong to states of the collision complex and exist only at short range, where the collision partners are close together.  This invites a treatment in terms of multichannel quantum defect theory (MQDT), which naturally divides the collision event into short-range physics where the resonances reside, and long-range physics that determines threshold behavior. As a result, we use the resonant spectrum to define a short-range K-matrix (here a $1 \times 1$ matrix in the single-open channel case we are considering), given by
\begin{equation}
\label{eq:K_mat}
    K^{\rm sr}(E) = -\pi \sum_{\mu} \frac{W_{\mu}^2}{E-E_\mu}.
\end{equation}
This is the quantity that is obtained from the difficult process of constructing a full potential energy surface and performing complete close-coupling calculations. Instead of carrying out this effort, we simulate many such calculations in our model by choosing random spectra of resonances.

Since $K^{\rm sr}$ is defined at short range, we can use the standard procedures of MQDT to construct relevant scattering observables, as described in detail in \cite{Ruzic_2013}. For these purposes, it is useful to represent quantities in scattering units for ultracold scattering: a natural length $\beta$, energy $E_\beta$, and time $\tau_\beta$ are given by
\begin{equation}
\label{eq:units}
\begin{split}
    \beta &= (2m C_6/\hbar^2)^{1/4} \\
    E_\beta&=\hbar^2/2m\beta^2 \\
    \tau_\beta &= 2m\beta^2/ \hbar,
\end{split}
\end{equation}
using the \response{reduced mass $m$}, van der Waals coefficient $C_6$, and reduced Planck's constant $\hbar$. Under MQDT, incorporating the effect of wave function propagation at long range then produces a full, energy-dependent $K$-matrix
\begin{equation}
\label{eq:sr_to_lr}
    K(E) = \frac{AK^{\text{sr}}(E)}{1 + \mathcal{G}K^{\text{sr}}(E)}
\end{equation}
{\final{in terms of the MQDT parameters $A$ and $\mathcal{G}$. Due to the ultracold temperatures under consideration, we assume that collisions occur in the Wigner threshold regime and are dominated by the $L=0$ partial wave, for which the MQDT parameters are given by
\begin{equation}
\label{eq:mqdt_params}
\begin{split}
    A &= \bar a k \\
    \mathcal{G} &= \left(\frac{\beta^2}{3}-\bar a^2\right)k^2 \\
    \bar a &= \left[\frac{\pi 2^{-3/2}}{\Gamma(5/4)\Gamma(1/2)}\right]^2 \beta,
\end{split}
\end{equation}
where $k$ is the scattering wave vector and ${\bar a}$ is the Gribakin-Flambaum characteristic scattering length \cite{gribakin_pra}.
From this procedure we obtain the scattering matrix}}
\begin{equation}
\label{eq:S_mat}
    S(E) = \frac{1+iK(E)}{1-iK(E)},
\end{equation}
which incorporates all the resonances in the model, as well as the effects of threshold scattering \footnote{{\final{In Eq.~(\ref{eq:S_mat}), we have omitted the smooth long-range reference phase that ordinarily multiplies the MQDT scattering matrix \cite{Ruzic_2013}. Its energy dependence contributes a background time delay associated with long-range propagation, rather than with residence in the short-range collision complex. Since our primary interest is the time delay associated with the short-range dynamics, we choose to not include such a term.}}}.

\subsection{Time Delay}
Using the scattering matrix, the lifetime of the collision complex can be evaluated using the Wigner-Smith time delay
\cite{wigner,eisenbud,smith}, given by
\begin{equation}
\label{eq:Q_mat}
    Q(E) = -i\hbar S^* \frac{\partial S}{\partial E}.
\end{equation}
\response{For resonant scattering, the time delay is inversely proportional to the resonance width and is associated with the lifetime of the corresponding quasibound state \cite{frye_2019}. We therefore use $Q(E)$ as a proxy for the lifetime of a collision complex formed at energy $E$. This identification is most direct when many resonances are present, because $Q(E)$ then reflects the lifetimes of the complex’s quasibound states. For nonresonant scattering, however, $Q(E)$ can contain substantial long-range contributions and therefore need not directly represent the short-range complex lifetime, as discussed further in Sec.~\ref{sec:sparse}. By contrast, in the dense-resonance case, sufficiently narrow resonances dominate the time delay, rendering long-range effects negligible. }



When discussing lifetimes, we must also take into consideration that collision energies in a real gas are distributed according to its thermal statistics. For a 3D classical gas with temperature $T$, collision energies are distributed according to a Maxwell-Boltzmann distribution
\begin{equation}
    P(E) = 2\sqrt{\frac{E}{\pi (k_B T)^3}} \exp\left(-\frac{E}{k_B T}\right).
\end{equation}
We therefore define the average time delay $\langle Q \rangle$ to be the thermally averaged value of $Q(E)$, resulting in
\begin{equation}
\label{eq:expected_Q}
    \langle Q \rangle \equiv \int_0^\infty dE\ P(E)Q(E).
\end{equation}

\subsection{Computational Details}
As  described earlier, our goal is to sample many different model spectra for a fixed set of parameters $d$, $x$, and $k_B T$. To do so, for each spectrum we generate  $400$ resonances with energies $\{E_\mu\}$ and couplings $\{W_\mu\}$ according to the distributions given in Eq.~(\ref{eq:wigner-dyson}) and Eq.~(\ref{eq:couplings}). \response{The collision threshold is then placed at the center of the spectrum and set to $E = 0$. The finite number of resonances is a sufficient approximation for the full spectrum because near threshold, the contribution of a distant resonance scales as $W_\mu^2/E_\mu$. Since $W_\mu^2 \sim d$, while $|E_\mu|$ increases with the resonance’s distance from threshold, the contribution of an individual resonance decreases with resonance index. Although the harmonic series does diverge, we find numerically that 400 resonances is sufficient to converge the resulting time delay.}
Given each spectrum, we  then numerically calculate  $Q(E)$ using Eq.~(\ref{eq:Q_mat}). By considering all our generated spectra, we collect statistics on $\langle Q\rangle$ for each different set of parameters ($d$, $x$, $k_B T$).

\section{\label{sec:results}Results}
\ptitle{methodology}

We find that a key factor determining the results of the model is whether the colliding molecules have access to many resonances, or few resonances. Hence, in this section, we distinguish cases with ``dense resonances'' ($d \ll k_BT$), when the mean level spacing is much less than the temperature, from those with ``sparse resonances'' ($d \gg k_BT$). Both cases appear to occur in experiments, and therefore we use realistic values for our parameters $(d,x,k_B T)$ corresponding to two actual experimental settings in order to illustrate these two regimes.

\subsection{\label{sec:dense}Dense Resonances}
\ptitle{describe results for dense resonances and how they seem to match rrkm}
To begin, we discuss the parameters ($d$, $x$, $k_B T$) we use to study the dense regime ($d \ll k_BT$). This regime is well characterized by the ${\rm RbCs} + {\rm RbCs}$ collisions observed in \cite{Gregory_2020}, with the temperature of the gas given by $T=2.2 $ $\mu$K. For this system, the mean level spacing is predicted to be $d = 0.2$ $\mu$K \cite{RbCs_DOS}, and the van der Waals energy is $E_{\beta} \approx 2.9$ $\mu$K \cite{C6}.  Given that $k_BT \approx E_\beta$ in this setting, we simply set $k_BT = E_{\beta} = 2.9\ {\rm \mu K}$ for convenience in the calculations that follow. In scattering units, the mean level spacing is given by $d=0.067\  E_{\beta}$, so this system is well within the dense regime $d \ll k_BT$. As mentioned earlier, we assume optimal bound-free coupling, $x=1$, in the results of this section. To summarize, this system is parametrized by 
\begin{equation}
\label{eq:dense_parameters}
\begin{split}
    d &=0.067\ E_\beta \\
    x &=1\\
    k_BT &=1\ E_\beta.
\end{split}
\end{equation}
Later in the section, we will remark on the effect of changing the value of these parameters while remaining in the dense regime.

\begin{figure}
\centering
\includegraphics[width=\linewidth]{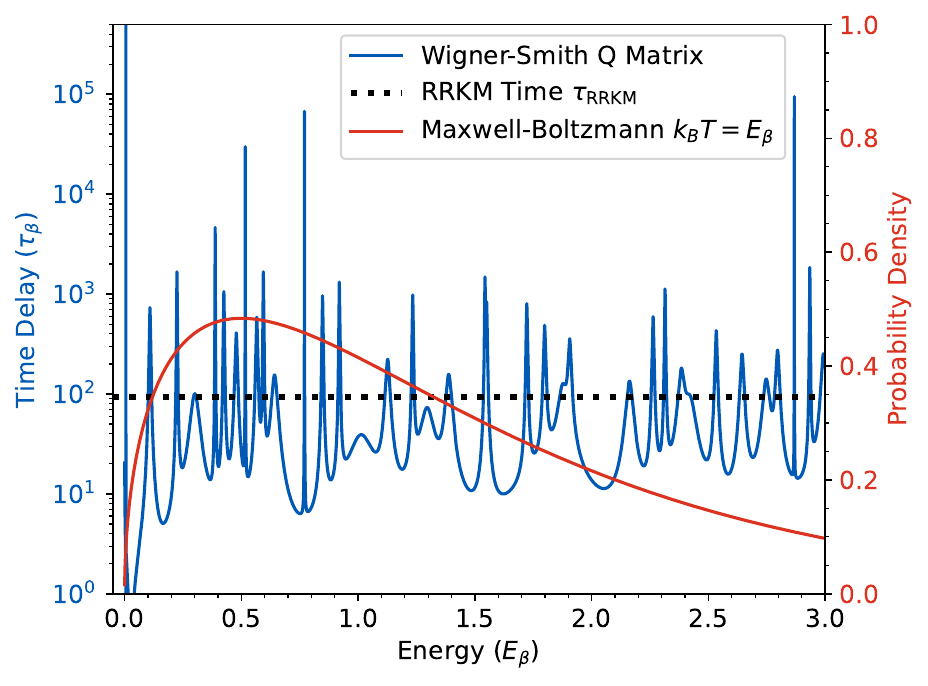} 
\caption{Wigner-Smith time delay $Q$ (solid blue) as a function of energy for one example spectrum in the dense regime, with the RRKM time (dotted black) shown for comparison. The Maxwell-Boltzmann distribution (red) is also plotted for $k_BT = E_\beta$.}
\label{fig:dense_spectrum}
\end{figure}

Figure \ref{fig:dense_spectrum} shows how the time delay $Q(E)$ varies as a function of energy for an exemplary spectrum with parameters from Eq.~(\ref{eq:dense_parameters}). As intended in our model, the structure of $Q(E)$ is characterized by its resonant features, and the overall structure is very similar in shape to the time delay derived from an actual close-coupling calculation \cite{Croft_2017}. \response{As noted previously, in this resonance-dominated regime, the Wigner–Smith time delay $Q$ can be associated directly with the lifetimes of quasibound complex states and therefore provides a meaningful proxy for the characteristic collision-complex lifetime.} To illustrate how many resonances contribute to collisions in the gas, we also plot the Maxwell-Boltzmann (MB) distribution of energies for $k_B T = E_{\beta}$. As expected in the dense regime, there are many resonances that span the temperature range, which means that many resonances will contribute when taking the thermal average. This is one of the key assumptions of RRKM theory, which predicts that averaging over many resonances should result in the lifetime of complexes being directly proportional to the density of states $\rho$. This argument produces the RRKM lifetime for one channel
\begin{equation}
\label{eq:rrkm}
   \rrkm = 2\pi\hbar \rho = \frac{2\pi\hbar}{d}.
\end{equation}
In the ${\rm RbCs} + {\rm RbCs}$ experiment, the RRKM lifetime is about $\rrkm = 93 \ \tau_{\beta} = 0.25\ {\rm ms}$, and we also plot this value (dotted black) in Fig.~\ref{fig:dense_spectrum}. From the figure, we can see that the RRKM time gives a reasonable scale for the average time delay, although $Q(E)$ can be much greater when on resonance. Overall, this suggests that $\rrkm$ is the correct theory in the dense regime, and we explore this connection more explicitly in the next section.

\subsubsection{Distribution of Time Delays}
\begin{figure}
    \centering
    \includegraphics[width=\linewidth]{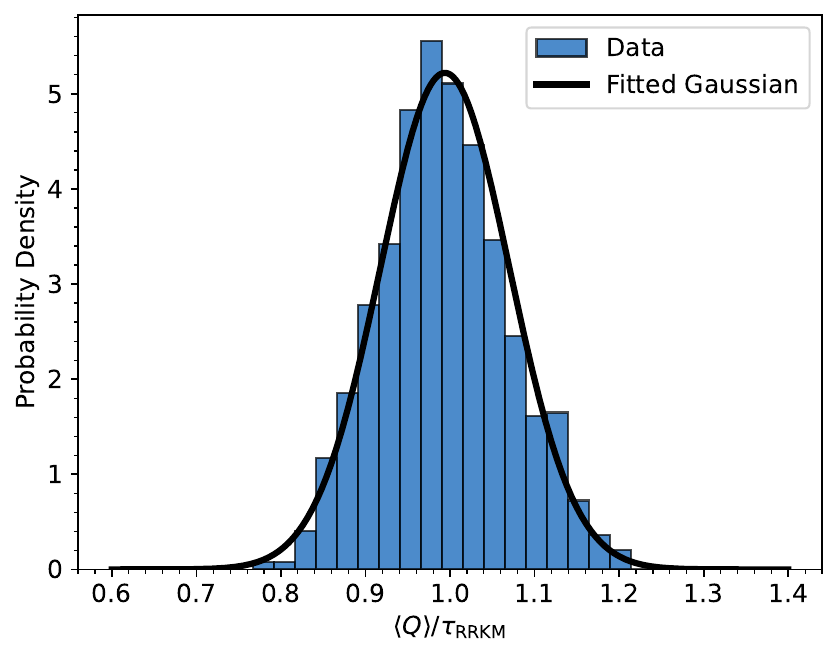}
    \caption{Expected time delay $\langle Q\rangle$ normalized by $\rrkm$ for 1000 randomly generated spectra with parameters from Eq.~(\ref{eq:dense_parameters}). The best-fit Gaussian is overlaid for comparison.}
    \label{fig:dense_thermal}
\end{figure}

Using our definition in Eq.~(\ref{eq:expected_Q}), we calculate the thermally averaged time delay $\langle Q \rangle$ for 1000 random spectra, in order to collect statistics on the possible lifetimes of the collision complex. Fig.~\ref{fig:dense_thermal} plots the  histogram of values for $\langle Q\rangle$, in units of $\rrkm$ for easy comparison.
As seen from the figure, the distribution is well centered around $\rrkm$ with Gaussian shape, with mean $\overline{\mu}_{\langle Q\rangle} = 0.99\ \rrkm$ and standard deviation $\sigma_{\langle Q\rangle} \approx 0.076\ \rrkm$. These statistics seem to imply that $\rrkm$ is in fact a very good estimate of the average value of $\langle Q \rangle$ in the model. 
This is because multiple resonances contribute to $\langle Q\rangle$ for any spectrum, which means the distribution of time delays will be fairly peaked around its mean ($\rrkm$). 

\begin{figure}
    \centering
    \includegraphics[width=\linewidth]{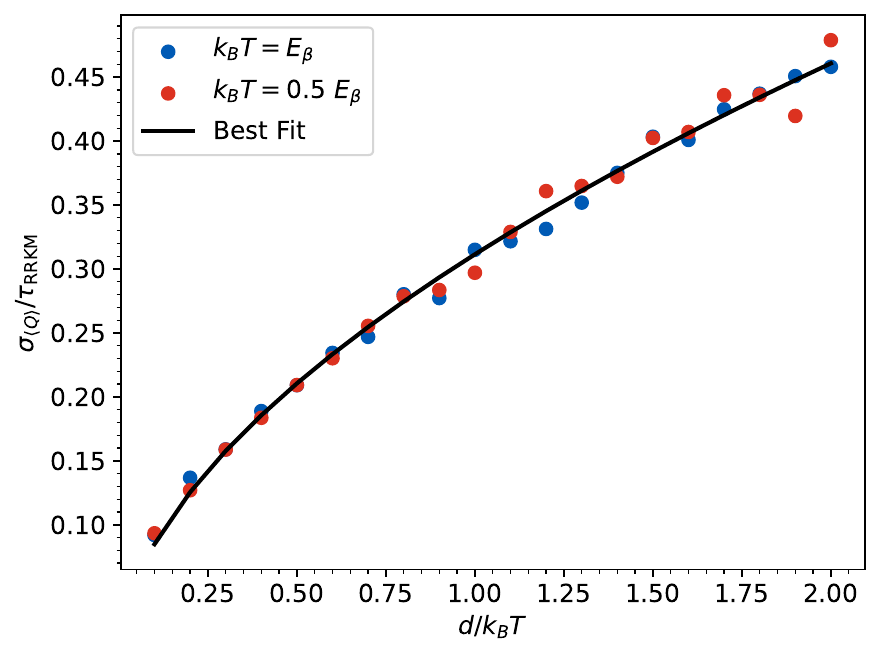}
    \caption{Dependence of $\sigma_{\langle Q\rangle}$ (in units of $\rrkm$) on the ratio $d/k_BT$ for $k_BT = E_\beta$ and $k_BT = 0.5\ E_\beta$. The line of best fit for a power law scaling $\sigma_{\langle Q\rangle} \sim (d/k_BT)^\alpha$ is also plotted as a solid line.}
    \label{fig:dense_sqrt}
\end{figure}

\begin{figure*}
\includegraphics[width=\linewidth]{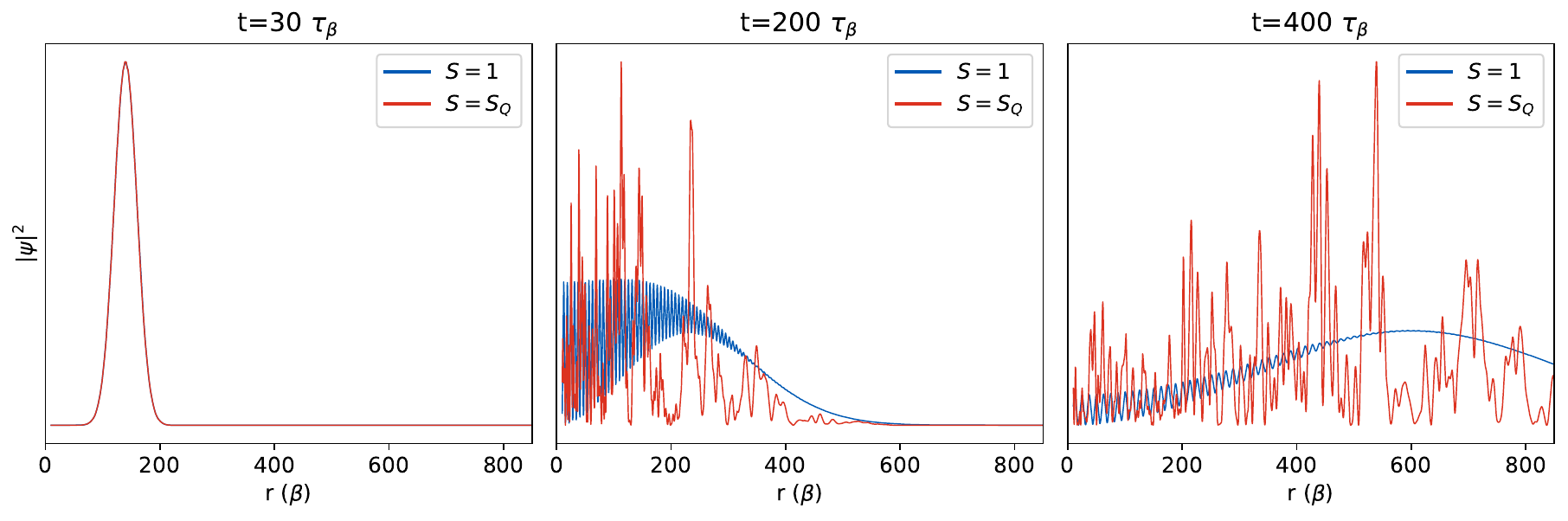} 
\caption{Evolution of a Gaussian wavepacket with $S=S_Q$ (resonant, red) and $S=1$ (trivial evolution, blue). The times correspond to the initial wavepacket ($t=30\  \tau_\beta$), the wavepacket interfering at short range ($t=200\  \tau_\beta$), and the wavepacket exiting from the collision ($t=400\  \tau_\beta$).}
\label{fig:evolution}
\end{figure*}

A qualitative explanation for this distribution can be seen as follows.  We assume a collection of $N$ isolated resonances within the energy range $k_BT$, such that $N \sim k_BT/d \gg1$ in the dense regime. Each isolated resonance has a contribution to the lifetime given by the resonant time delay operator defined above.  In the optimal coupling case $x=1$, and ignoring the MQDT parameter ${\cal G}$ in the low-energy limit, the procedure above produces the usual Lorentzian time delay for the resonance at $E_\mu$, given by
\begin{align}
Q_\mu(E) \approx \frac{ \hbar \gamma_\mu }{ (E - E_{\mu})^2 + (\gamma_\mu/2)^2 },
\end{align}
with the linewidth {\final{$\gamma_\mu = 2 \pi {\bar a}k W_{\mu}^2$}}. \final{On average, neighboring resonances lie approximately \(d\) away on either side of \(E_\mu\). Assigning each resonance an energy interval bounded halfway to its neighbors, the averaged contribution is}
\begin{equation}
\begin{split}
    \langle Q_\mu \rangle &= \frac{ 1 }{ d } \int_{E_{\mu} - d/2}^{E_{\mu}+d/2}
dE\ \frac{ \hbar \gamma_\mu }{ (E - E_{\mu})^2 + (\gamma_\mu/2)^2 } \\
&= \frac{ 4 \hbar }{ d } \tan^{-1} \left( \frac{ d }{ \gamma_\mu } \right)
\end{split}
\end{equation}
Using the mean value $W_\mu^2 = \langle W_{\mu}^2 \rangle = d/\pi^2$, this becomes essentially independent of the specific resonance, so we can define the value
\begin{align}
\langle Q \rangle_0 \equiv \frac{ 4 \hbar }{ d } \tan^{-1} \left( \frac{ \pi }{ 2 {\bar a} k } \right).
\end{align}
Assuming low enough energies where \final{$\pi / 2{\bar a}k \gg 1$}, the arctangent is approximately $\pi/2$, such that the averaged contribution is
\begin{align}
\langle Q \rangle_0 = \frac{ 2 \pi \hbar }{ d } = \rrkm.
\end{align}
In the regime $d \ll k_BT$, thermal averaging only picks out the averaged contribution from each resonance because the width of the MB distribution is much larger than the resonance widths. All of these averaged contributions will contribute $\rrkm$ such that the total average is also $\rrkm$. Equivalently, we can write
\begin{equation}
\begin{split}
    \langle Q \rangle &= \int_{0}^\infty dE\ P(E)Q(E) \\
    &\approx \sum_{\mu} (E_\mu-E_{\mu-1})P(E_\mu)\langle Q_\mu\rangle \\
    &\approx \langle Q\rangle_0 \approx \rrkm.
\end{split}
\end{equation}
This explains why the mean of the distribution in Fig.~\ref{fig:dense_thermal} is  centered at $\rrkm$. Notably, this result is analogous to what was found in \cite{frye_2019, transition_state_theory}, where the contribution to a thermally averaged time delay is independent of the resonance width. 

\subsubsection{Distribution Width}

The ability to run simulations with many different sample spectra yields information about possible distributions of time delays, not merely single estimated values.
Qualitatively, we can make a prediction of the width of the $\langle Q\rangle$ distribution by using counting statistics. Each resonance contributes a time delay $\langle Q_\mu\rangle$, where $\mu=1, \dots, N$ and $N = k_BT/d$ is the total number of contributing resonances, approximately the number within $k_BT$.  These $N$ values are random variables with mean $\rrkm$.  Invoking the central limit theorem for large $N$, one would then expect a Gaussian distribution with mean $\rrkm$ and a standard deviation satisfying
$\sigma_{\langle Q \rangle} / \mu_{\langle Q \rangle} \sim 1/\sqrt{N} \sim 1/\sqrt{ k_BT/d}$, whereby the width of the distribution is given by
\begin{align}
\label{eq:sqrt}
\sigma_{\langle Q \rangle} \sim \tau_{\rm RRKM} \sqrt{ \frac{ d }{ k_BT } }.
\end{align}

To confirm this, we compute $\sigma_{\langle Q\rangle}$ for various ratios of $d/k_BT$. In Fig.~\ref{fig:dense_sqrt}, we vary the value of $d$ for two temperatures $k_B T = E_\beta$ and $k_B T = 0.5\ E_\beta$ and plot $\sigma_{\langle Q\rangle}$ as a function of the resulting ratio $d/k_B T$. Fitting the data to a power law scaling gives 
\begin{equation}
    \frac{\sigma_{\langle Q \rangle}}{\rrkm} = (0.31 \pm 0.002) \times \left(\frac{d}{k_BT} \right)^{0.565\pm 0.01}.
\end{equation}
\response{The fitted exponent exhibits a small deviation from $1/2$, which may be due to breakdown of our assumption of resonances as independent random variables.} Nonetheless, this simple argument gives qualitatively correct predictions for behavior in the dense regime: changing the mean level spacing $d$ or the temperature $T$ in this regime will only change the resulting width of the distribution based on how many resonances are being averaged over. In Appendix \ref{ap:coupling}, we elaborate on the effect of changing the last parameter, $x$.



\subsubsection{Wavepackets}

The time delay $Q(E)$ is evaluated in the energy domain, but it can also be instructive to note how the resonant spectrum explicitly influences the propagation of wavepackets in the time domain.
Returning to the initial parameters in Eq.~(\ref{eq:dense_parameters}), we use the resonant spectrum corresponding to Fig.~\ref{fig:dense_spectrum} to propagate a Gaussian wavepacket through the scattering process.  Details of the wavepacket propagation are given in Appendix \ref{ap:wavepacket}; here we present only the results. In Fig.~\ref{fig:evolution}, we show the evolution of a Gaussian wavepacket with scattering matrix $S=S_Q$, where $S_Q$ is the scattering matrix corresponding to the resonances in Fig.~\ref{fig:dense_spectrum}. As a baseline, we also plot the evolution of the same initial wavepacket under trivial evolution, which is given by $S=1$ such that there are no resonances. The first feature of note is that the wavepacket under the influence of the resonances (red) becomes ``jagged'' compared to the wavepacket with no resonances (blue) after the collision at $r = 0$. This is due to the fact that the resonances in $S_Q$ result in rapidly varying phase shifts, which destroy the phase coherence of the original Gaussian wavepacket and cause the outgoing wavepacket to be made of momentum components with essentially random phases. 

\begin{figure}
    \centering
    \includegraphics[width=\linewidth]{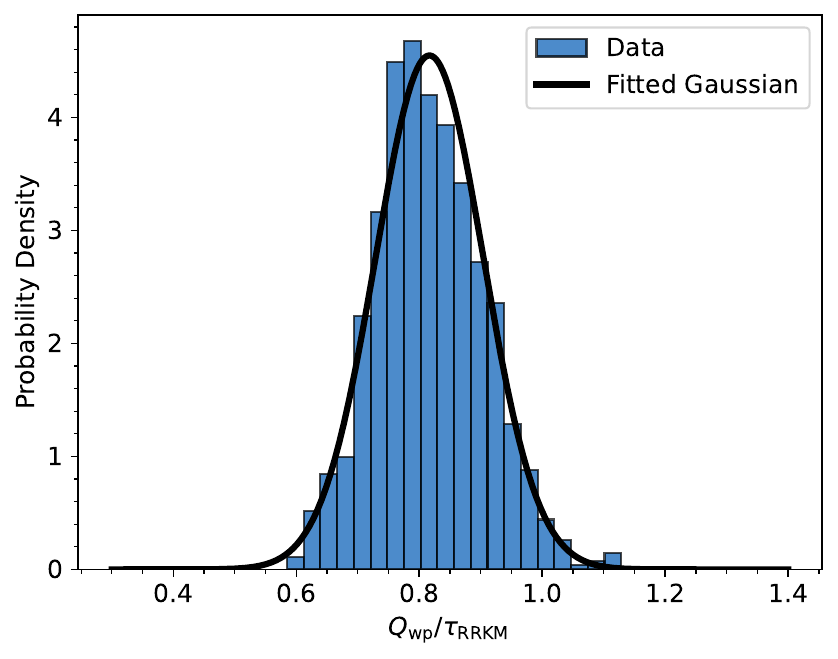}
    \caption{Wavepacket time delay $Q_{\rm wp}$ from wavepacket evolution for 1000 randomly generated spectra with parameters from Eq.~(\ref{eq:dense_parameters}). The best-fit Gaussian is overlaid for comparison.}
    \label{fig:dense_wavepacket}
\end{figure}

We also observe that the resonant wavepacket seems to be \textit{delayed} in comparison to the trivially evolving wavepacket after the collision: the ``center of mass'' of the resonant wavepacket trails behind the center of the other wavepacket. This is in fact the physical effect of the time delay $Q$, and we associate this delay with the sticking time of the complex. As can be seen in Fig.~\ref{fig:evolution}, however, there is not a uniform delay across the wavepacket. Instead, the probability density leaks out slowly from the collision, which means that the collision complex lifetime should be treated as the mean of a distribution of exit times.

Additionally, we extract a wavepacket time delay $Q_{\rm wp}$ by comparing the average positions of wavepackets with and without resonances (see Appendix \ref{ap:wp_time_delay} for details).
Using this method, we compute the distribution of wavepacket time delays $Q_{\rm wp}$ for 1000 random spectra, plotted in Fig.~\ref{fig:dense_wavepacket}. Examining the distribution, the mean $\overline{\mu}_{\rm wp} = 0.82\ \rrkm$ is less than the RRKM time, although the shape remains Gaussian with small standard deviation $\sigma_{\rm wp} = 0.087\ \rrkm$. The discrepancy here can be explained by the narrow resonances that we can observe in Fig.~\ref{fig:dense_spectrum}. These narrow resonances have very large peaks spanning only a small energy range, which results in very long time delays for a small portion of the wavepacket. The numerical calculation, however, is limited by finite energy resolution and grid size, causing an underestimation of the true time delay \response{(see Appendix \ref{ap:wavepacket} for details)}. Nevertheless, the overall error is within $20\%$ of $\rrkm$, consistent with the picture of RRKM being the correct theory in the dense regime.

\subsection{Transition from Dense to Sparse Resonances}

\begin{figure}
    \centering
    \includegraphics[width=\linewidth]{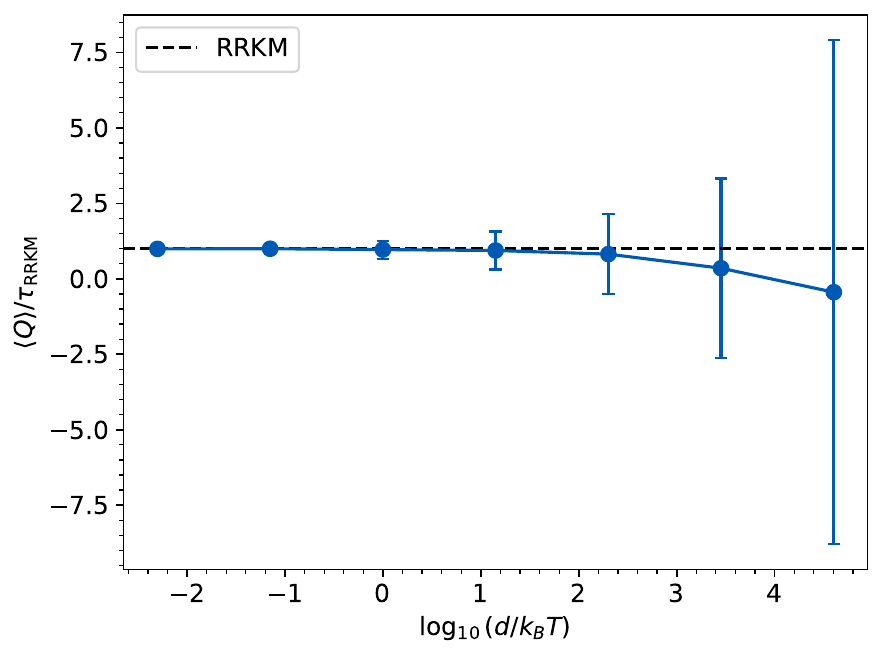}
    \caption{Mean (points) and standard deviation (error bars) of time delay $\langle Q \rangle$ normalized by $\rrkm$ as a function of $d$ when $k_BT = E_\beta$. For each value of $d$, the statistics were computed over 1000 random spectra. The dotted line represents $\langle Q \rangle = \rrkm$, showing breakdown of RRKM scaling as $d/k_BT$ increases. }
    \label{fig:mean_spacing_delays}
\end{figure}

In Fig.~\ref{fig:dense_sqrt}, we studied how the time delay changes when varying the mean level spacing $d$ within the dense regime. As a natural continuation, we investigate how the behavior of the time delay changes as we exit the dense regime characterized by many contributing resonances. Still retaining the parameters $x=1$ and $k_B T = 1\ E_\beta$, we plot in Fig.~\ref{fig:mean_spacing_delays}  the mean and standard deviation of $\langle Q \rangle / \rrkm$ as we vary the value of $d$ from \final{$10^{-2} E_\beta$ to $10^4 E_\beta$}. For small values of $d / k_B T$, the mean remains at \smash{$\overline{\mu}_{\langle Q\rangle} = \rrkm$}, while the standard deviation increases with $d$ due to fewer  resonances being averaged over, consistent with our qualitative expectation in Eq.~(\ref{eq:sqrt}). 
In the other limit, we see that when $d \gg k_B T$, the mean of $\langle Q\rangle$ drops to $0$, while the standard deviation is now so large that a wide variety of time delays, many negative, are now possible. The features of this new distribution are discussed in the following subsection.


\begin{figure}
    \centering
    \includegraphics[width=\linewidth]{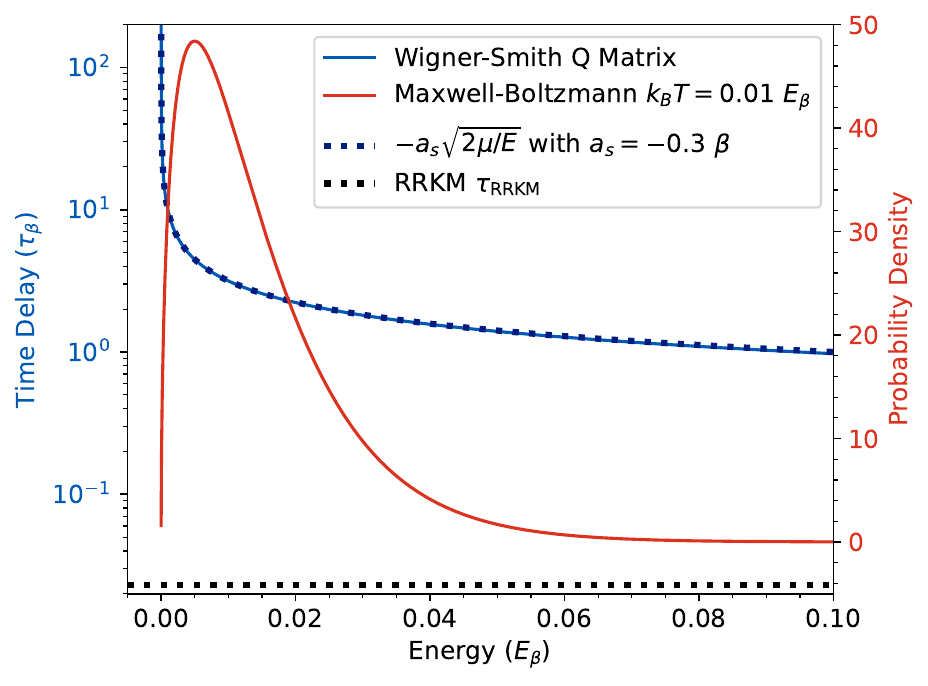}
    \caption{Wigner-Smith time delay $Q$ (blue) as a function of energy for one example spectrum in the sparse regime, with the RRKM time (dotted black) shown for comparison. The curve of time delay corresponding to the fitted scattering length of $a_s=-0.3\ \beta$ (dotted blue) is also plotted, demonstrating agreement with threshold scaling in Eq.~(\ref{eq:threshold_Q}). Finally, the Maxwell-Boltzmann distribution (red) is plotted for $k_BT = 0.01\ E_\beta$.}
    \label{fig:sparse_spectrum}
\end{figure}

\subsection{\label{sec:sparse}Sparse Resonances}
\ptitle{postulate what is going on for sparse resonances}
Owing to the extremely low temperatures in ultracold gas experiments, a common situation for alkali-alkali dimer collisions is that they are in the sparse limit ($d \gg k_BT$).  The characteristic system we consider for the sparse regime is given by the ${\rm Rb} + {\rm KRb}$ collisions observed at $480$ nK in \cite{nichols_PhysRevX.12.011049}. We take the mean level spacing (for ro-vibrational states, ignoring the possibility of spin degrees of freedom) to be $d=12$ mK (corresponding to $\rrkm \approx 4$ ns), far greater than the temperature of the gas in the experiment.
In our scattering units, $E_\beta = 45\ {\rm \mu K}$, so the parameters we use to investigate the limit of sparse resonances are thus initially given by 
\begin{equation}
\label{eq:sparse_parameters}
\begin{split}
    d&=270\ E_\beta \\
    x&=1\\
    k_BT &=0.01\ E_\beta.
\end{split}
\end{equation}

Fig.~\ref{fig:sparse_spectrum} shows a sample spectrum in this regime, along with the Maxwell-Boltzmann distribution for $k_BT = 0.01\ E_\beta$. Comparing with the dense case in Fig.~\ref{fig:dense_spectrum},  the lack of resonances is immediately apparent (and unsurprising given that $d \gg k_B T$).
Instead of resonant structure, $Q(E)$ smoothly diverges as the energy tends to 0. 
This is because in the Wigner threshold regime ($E \ll E_\beta$), the $S$ matrix  is given by $S(E)=\exp(-2ika_s)$ with $k = (2m E/\hbar^2)^{1/2}$ and $a_s$ the scattering length. The time delay resulting from this $S$ matrix is then given by
\begin{equation}
\label{eq:threshold_Q}
    Q(E) = - a_s\sqrt{\frac{2m}{E}}
\end{equation}
according to Eq.~(\ref{eq:Q_mat}). For comparison, the dotted line in  Fig.~\ref{fig:sparse_spectrum} plots Eq.~(\ref{eq:threshold_Q}) using the scattering length $a_s = -0.3\ \beta = -62\ a_0$ (obtained from best fit), and we see very good agreement with our prediction for the scaling of $Q(E)$. 

Given that the time delay in the threshold regime is characterized by a single scattering length, one might wonder whether the properties of the resonant spectrum contribute at all to the lifetime. It should be noted, however, that if there were truly no resonances in the spectrum, Eq.~(\ref{eq:K_mat}) would imply that the $S$-matrix would be energy-independent, with zero time delay. Rather, in the sparse regime, we can interpret the resonances as collectively contributing to the value of the scattering length $a_s$. Specifically, each resonance contributes some off resonant scattering phase shift in the threshold limit. These phase shifts collectively contribute to a total short-range phase shift \cite{julienne_background}, which determines the value of $a_s$ when combined with the contribution from the long-range potential. Investigating the influence of the resonant spectrum on the time delay in this regime therefore amounts to investigating the distribution of scattering lengths.

An important subtlety is that, so far, we have treated the time delay $Q(E)$ as an indicator of the complex lifetime. \response{At threshold, however, $Q(E)$ is strongly influenced by the long-range potential and therefore includes long-range dwell time and effects such as quantum reflection. In addition, the long-range interaction can support bound states that couple only weakly to the chaotic short-range complex \cite{croft26_longrange}, a mechanism not included in our model. These effects make it difficult to interpret the time delay in the sparse, threshold-dominated regime as a short-range complex lifetime. This is in contrast to the dense regime, where resonant delays can be associated more directly with quasibound states of the complex. }

\response{For this reason, in the sparse limit we interpret the time delay as a quantity tied to the scattering length, which incorporates long-range effects. From this perspective, the time delay should be regarded as a generic feature of the scattering process and should not, in this limit, be conflated with the lifetime of the complex. More broadly, the main conclusion we emphasize is that the RRKM estimate, which is based on averaging over many resonances, is not an appropriate reference scale in the sparse limit.}

\subsubsection{Distribution of Time Delays}

\response{Despite the influence of the long-range potential, the distribution of time delays and scattering lengths can still yield valuable insights in the sparse limit.} 
Using Eq.~(\ref{eq:threshold_Q}), the thermally averaged time delay is given by
\begin{equation}
\begin{split}
    \langle Q \rangle &= \int_{0}^\infty dE\ P(E) \frac{-a_s\sqrt{2m}}{\sqrt{E}} \\
    &= -2a_s\sqrt{\frac{2m}{\pi k_B T}}.
\end{split}
\end{equation}
As we mentioned, this quantity may vary widely with the different scattering lengths $a_s$ in each realization of the spectrum.  However, a typical characteristic scale for $\langle Q \rangle$ is set by replacing $a_s$ with the standard Gribakin-Flambaum scattering length ${\bar a}$ \cite{gribakin_pra}.
The scattering unit in which to express time delays in the sparse limit is therefore the characteristic delay time  $\tau_T = 2\bar a \sqrt{2m/\pi k_B T}$. Significantly, this characteristic time is independent of the mean level spacing $d$ and hence fundamentally different from the RRKM time. This reflects the distinct physics of the sparse regime, where too few resonances lie within the thermal window for the resonance averaging assumed by RRKM theory.

\begin{figure}
\begin{tabular}{c}
\includegraphics[width=\linewidth]{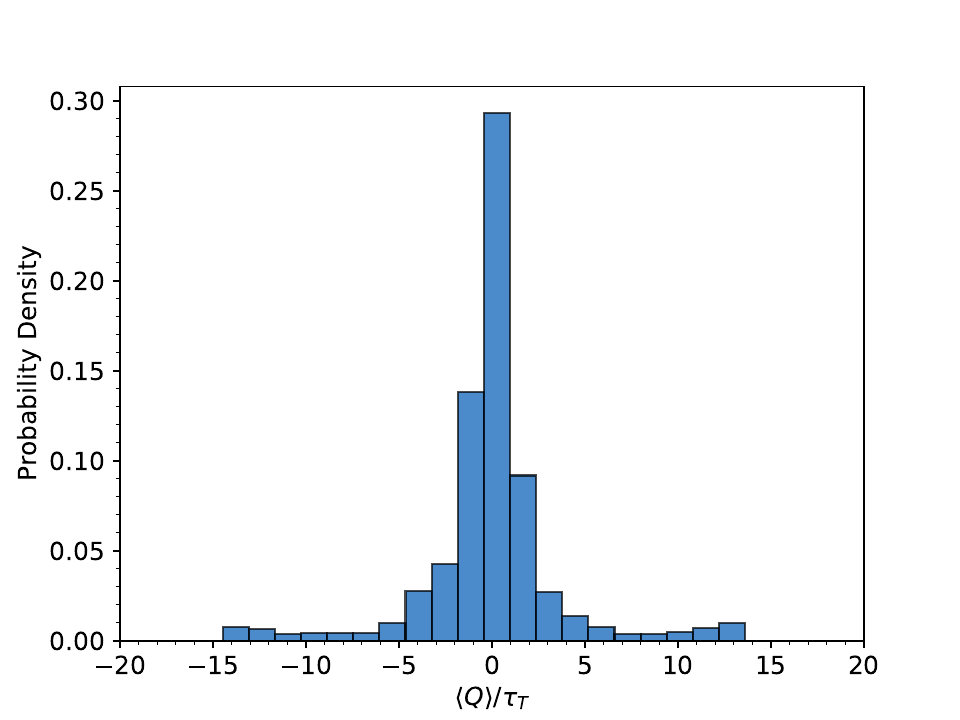} \\
(a) \\
\includegraphics[width=\linewidth]{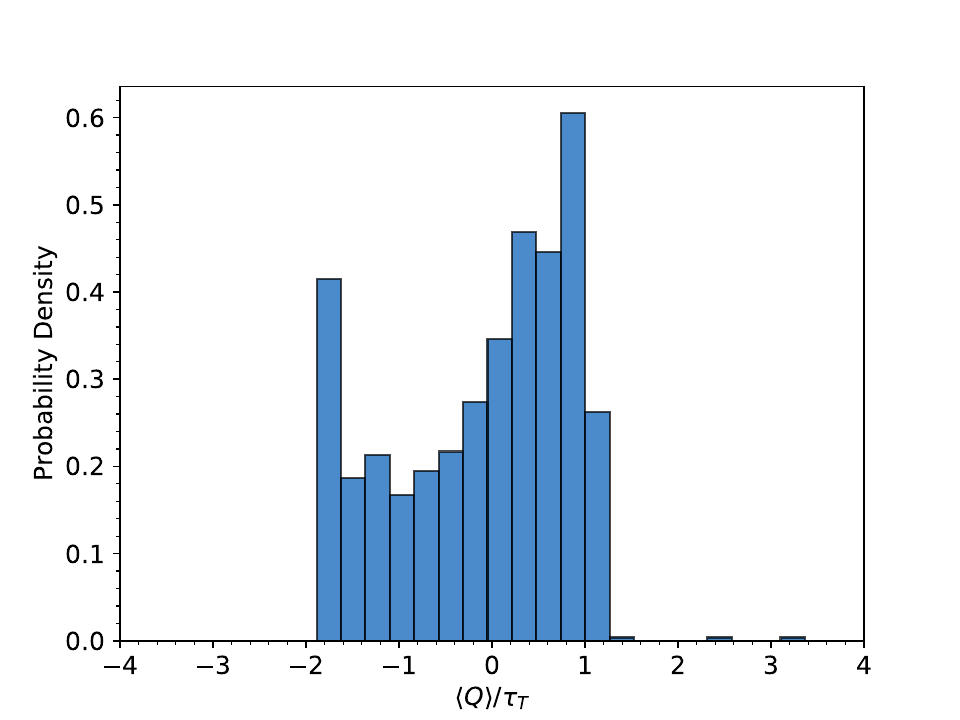} \\
(b)
\end{tabular}
\caption{Expected time delay $\langle Q\rangle / \tau_T$  for 1000 randomly generated spectra for (a) $k_BT = 0.01\ E_\beta$ and (b) $k_BT = 1\ E_\beta$. In both, we fix the parameters $d=270\ E_\beta$ and $x=1$. For the ${\rm Rb} + {\rm KRb}$ collisions, $\tau_T$ is \final{about $900\ {\rm ns}$ in (a) and $90\ {\rm ns}$ in (b).}}
\label{fig:sparse_thermal}
\end{figure}

We plot in Fig.~\ref{fig:sparse_thermal}(a) the histogram for 1000 values of $\langle Q \rangle / \tau_T$ for the initial parameters in Eq.~(\ref{eq:sparse_parameters}), and we observe several differences from the dense regime. Firstly, the shape is clearly non-Gaussian, as opposed to the dense case shown in
Fig.~\ref{fig:dense_thermal}. 
Secondly, the scale of the distribution is correctly given by $\tau_T$: for the ${\rm Rb} + {\rm KRb}$ collisions, \final{$\tau_T \approx 900\ {\rm ns}$} when $k_B T = 0.01\ E_\beta$ as in the experiment. For comparison, we recall that the RRKM time is about $4\ {\rm ns}$, so most values of $|\langle Q \rangle|$ are several orders of magnitude greater than $\rrkm$. 
Finally, we see that the distributions of time delay now have a significant probability for $\langle Q \rangle$ to be negative. In the threshold picture, this is a result of a positive scattering length,
corresponding to quantum reflection at long range before the complex is formed.

From Eq.~(\ref{eq:threshold_Q}), it is clear that the time delay should depend on temperature.  In Fig.~\ref{fig:sparse_thermal}(b) we therefore plot the same histogram, but at $k_BT = 1\ E_\beta$.
From these two histograms, the shape of the distribution of time delays is shown to be strongly temperature dependent in the sparse regime. Interestingly, simulations also show that the distributions in Fig.~\ref{fig:sparse_thermal} are completely independent of mean resonance spacing $d$ as long as we remain in the limit of sparse resonances---that is, the scaled distributions look identical regardless of the value of $d$. As a result, the two regimes are in some sense opposites of each other: in the dense regime, the time delay is characterized by $d$ and not $k_BT$, whereas in the sparse regime, $k_BT$ is the relevant parameter and $d$ has no influence.

As we mentioned before, the time delays in the sparse regime correspond to the scattering length of the collisions. However, the distributions in Fig.~\ref{fig:sparse_thermal} differ significantly from the distributions of scattering lengths predicted for general potential scattering in \cite{wang_bohn}. This is due to the fact that we have imposed specific resonant structure in the model using Eq.~(\ref{eq:K_mat}), so our results describe the scattering lengths resulting from resonances obeying random matrix theory, rather than general potentials. In order to understand this distribution better, as well as the reason that the mean level spacing is  unimportant in this regime, we examine a simplified one resonance model in the next section.


\subsubsection{One Resonance Model}

In the sparse regime, we can understand the characteristic properties by considering a simplified model of resonances. Noting that $d \gg k_BT$ and that the contribution from a resonance scales inversely to its distance since $K\sim (E-E_\mu)^{-1}$, we assume, for simplicity, that only the nearest resonance (to threshold $E=0$) influences the scattering length and hence the time delay. We label the energy and coupling of this closest resonance to be $E_0$ and $W_0$ respectively, and we let the detuning from collision energy $E$ be $\Delta_0 \equiv E_0-E$. 
A few numerical tests reveal that when the resonance energies are Wigner-Dyson distributed according to Eq.~(\ref{eq:wigner-dyson}), then $\Delta_0$ is approximately normally distributed with zero mean and \response{variance $0.16d^2$} (for explanation, see Appendix \ref{ap:wigner}). We also recall that the coupling $W_0$ is drawn from the Gaussian distribution as defined in Eq.~(\ref{eq:couplings}). As such, we can treat $\Delta_0$ and $W_0$ as Gaussian random variables 
\begin{equation}
\label{eq:distribution}
\begin{split}
    \Delta_0 &\sim N(0, 0.16d^2) \\
    W_0 &\sim N(0, xd/\pi^2).
\end{split}
\end{equation}

For a given value of $E_0$ and $W_0$, the time delay $Q(E)$ for one resonance can then be calculated analytically using the equations from Sec.~\ref{sec:theory} to yield
\begin{equation}
\label{eq:one_res}
    Q(E) =  \pi \bar a W_0^2 \sqrt{\frac{2m}{E}} \frac{(2+\zeta W_0^2)E+\Delta_0}{(\zeta W_0^2E - \Delta_0)^2 + \eta W_0^4 E},
\end{equation}
with constants (in scattering units)
\begin{equation}
\begin{split}
    \zeta &= 2m\pi(\bar a^2-\beta^2/3)/\hbar^2 = -0.33\ E_\beta^{-1} \\
    \response{\eta} &= \response{2m\pi^2\overline{a}^2/\hbar^2 = 2.25\ E_\beta^{-1}}.
\end{split}
\end{equation}
Since we are in the regime $d \gg E_\beta$, both $W_0^2$ and $E_0$ are likely to be much larger than $E_\beta$. Making these assumptions, 
we can write the time delay as approximately
\begin{equation}
\label{eq:invariant_Q}
\begin{split}
    Q(E) =& \pi \alpha_0 \bar a \sqrt{\frac{2m}{E}} \frac{\zeta E \alpha_0 + 1}{(\zeta E \alpha_0-1)^2+\eta E \alpha_0^2} \\
    \prescript{E\to0}{}{=}& \pi \alpha_0 \bar a \sqrt{\frac{2m}{E}},
\end{split}
\end{equation}
where we have defined $\alpha_0 \equiv W_0^2/\Delta_0$. Although the distributions of $\Delta_0$ and $W_0$ clearly depend on $d$, we see that $Q(E)$ only depends on the ratio $\alpha_0$, whose distribution is independent of $d$. This is because
the transformation $d \to cd$ takes $\Delta_0 \to c\Delta_0$ and $W_0 \to \sqrt{c} W_0$ in Eq.~(\ref{eq:distribution}).
As a result, the time delay in the sparse limit is expected to depend only weakly, if at all, on the mean level spacing, as we observed earlier.

While the characteristic time $\tau_T$ correctly sets a scale for the distribution of time delays in the sparse limit, we can see from Fig.~\ref{fig:sparse_thermal} that the shapes of these distributions are somewhat elaborate. From Eq.~(\ref{eq:invariant_Q}), the distribution of $Q(E)$ can  be understood using the distribution of $\alpha_0$. \response{Appendix \ref{ap:sparse} derives an approximate analytical expression for the distribution of this ratio, although the resulting form offers limited additional physical insight.} Here, however, we note one such outcome of the analysis, namely, that varying the dimensionless coupling constant $x$ can also influence the time delay. The analysis in Appendix \ref{ap:sparse} shows that the characteristic width of the distribution for $\alpha_0$ is about $0.18x$. In the threshold limit, a value of $|\alpha_0| = 0.18x$ corresponds to a characteristic scattering length of $|a_s|=0.58x\overline{a}$ and a characteristic time delay 
\begin{equation}
\label{eq:modified_sparse_Q}
    \response{Q_{\rm char}} = 0.65x\overline{a} \sqrt{\frac{2m}{k_B T}}
\end{equation}
For $x=1$, the resulting values give good agreement with the histogram in Fig.~\ref{fig:sparse_thermal}(a). 

The key point from this analysis is that the off-resonant contribution from the spectrum is controlled by the bound-free coupling parameter. In our model, the magnitude of the scattering length (and thus time delay) can be made arbitrarily large or small by tuning the value of $x$. While we postulate that most systems should have a value close to $x = 1$, it is not impossible that certain systems deviate significantly from unity, resulting in abnormally large scattering lengths and long time delays.

\subsubsection{Wavepackets}
We briefly remark on how the sparse regime translates to wavepacket evolution. Unlike the dense case, there are no resonances within the energy window, so $Q(E)$ varies relatively slowly, which causes an overall delay in the entire wavepacket, rather than creating many sharp features like in Fig.~\ref{fig:dense_wavepacket}. As a result, there is a uniform notion of time delay for any collision. Calculating $Q_{\rm wp}$ as we did above, we find again very similar distributions to Fig.~\ref{fig:sparse_thermal}.




\section{{Implications For} Experimental Measurements}
\label{sec:experiments}

To characterize the dense and sparse regimes, we chose model parameters $(d,x,k_BT)$ consistent with actual experimental systems. As a natural conclusion, we remark on how our results compare with the collision complex lifetimes measured by the experiments.

Representing the dense regime, the measured lifetime of ${\rm RbCs}+{\rm RbCs}$ collisions is about 0.53 ms, only about a factor of $2$ greater than the RRKM lifetime \cite{Gregory_2020}. This bodes well for the approximate success of our statistical model.  However, our oversimplified model still falls somewhat short of quantitative agreement with the measured lifetime. As seen in Fig.~\ref{fig:dense_thermal}, the standard deviation of $\langle Q \rangle$ around the mean $\rrkm$ is predicted to be less than $0.1\ \rrkm$. Therefore, the experimental measurement represents a lifetime which is more than 10 standard deviations from our prediction. This discrepancy could arise from a number of different reasons, such as an incorrect counting of the density of states or additional physics. Nonetheless,  given that the system should be well within the dense regime, we believe RRKM theory should ultimately be applicable for this system in some capacity. Notably, in this limit it would apparently be inappropriate to include the numerous nuclear spin degrees of freedom, as they would increase the RRKM lifetime by orders of magnitude, spoiling this apparent near-agreement.

In the sparse regime, on the other hand, the collision complex of ${\rm KRb}+{\rm Rb}$ collisions was measured to have a lifetime of about 0.39 ms, greater than $\rrkm$ by five orders of magnitude \cite{nichols_PhysRevX.12.011049}. As discussed earlier, however, the RRKM
lifetime has very little meaning in the sparse regime due to the absence of resonances. Rather, our model predicts a threshold time delay corresponding to a characteristic scattering length, given in Eq.~(\ref{eq:modified_sparse_Q}). In the case of ${\rm KRb}+{\rm Rb}$, this is given by a scattering length of $a_s = x(54\ a_0)$ and a time delay of $\langle Q \rangle = x(527\ {\rm ns})$. \response{As discussed in Sec.~\ref{sec:sparse}, however, the time delay in the sparse regime includes significant contributions from the long-range potential and does not directly translate to the lifetime of the short-range complex. Nevertheless, even assuming that the time delay is due solely to the complex, our results would \final{roughly require a dimensionless bound-free coupling parameter of $x=740$ and a scattering length of $4\times10^4\ a_0$ to match the measured lifetime.}}  While this is of course not impossible, it is a very particular requirement with no {\it a priori} justification. Likewise, it is also possible that there exists a narrow resonance at threshold that contributes to anomalously long time delays. However, experimental measurements have observed similarly long lifetimes while changing the hyperfine state and magnetic field \cite{KKN}; since tuning either presumably removes such a resonance, this explanation is also unlikely. 

Additionally, we note that the paucity of resonances is also seen in a very recent ${\rm KRb}+{\rm Rb}$ close-coupling calculation  \cite{kendrick_2026}, which does not find resonances below $1\ {\rm mK}$ even when including the effects of the conical intersection. The calculation does not, however, include spin degrees of freedom, which could introduce additional resonant features. Nevertheless, we argue that even including the effects of spin cannot populate a vast forest of resonances at the temperature of the experiment. As acknowledged in \cite{kendrick_2026}, nuclear spin degrees of freedom would amount to an increase in the density of states by a factor of $(2\times3/2+1)^2(2\times4+1)=144$ (the nuclear spins being $3/2,3/2,4$ for the three atoms). This would reduce the mean level spacing to around 90 $\mu$K, which is still not dense enough for the RRKM limit to apply.  Based on our results in Eq.~(\ref{eq:modified_sparse_Q}), this would not change the predicted lifetime.

\response{Finally, we consider the ${\rm NaK}+{\rm NaK}$ and ${\rm NaRb}+{\rm NaRb}$ collisions in \cite{Gersema_2021, bause_prr}, for which the measured complex lifetimes (lower bounds in the case of \cite{Gersema_2021}) exceed the RRKM predictions by roughly two orders of magnitude. In these experiments, however, $d/k_B T \approx 10$, placing both systems in the intermediate regime of Fig.~\ref{fig:mean_spacing_delays} rather than clearly in either the dense or sparse limit. The number of resonances contributing to the time delay may therefore depend sensitively on their precise positions. The discrepancy of two orders of magnitude could thus arise from an anomalously narrow resonance near threshold, an increased density of states due to nuclear-spin degrees of freedom, or other effects. Because these systems do not lie in either asymptotic regime, it is difficult to support a definitive conclusion for these systems.}


\section{Discussion}
\label{sec:conclusion}

We have studied how simulations of close-coupling calculations can be used to reason about collision complex lifetimes. By generating random resonant spectra according to random matrix theory, we were able to collect statistics of reasonable complex lifetimes for systems characterized by three parameters: mean level spacing $d$, bound-free coupling $x$, and temperature $k_BT$. Notably, we have shown that our statistical model shows entirely different physics when the system is in the dense regime ($d \ll k_BT$) versus the sparse regime ($d \gg k_BT$).


In the dense regime of many resonances, we have asserted that the RRKM lifetime is in fact a reasonable quantity, because the assumption of averaging over many resonances is satisfied. While we found that the deviation of time delays scales with the temperature, our model predicts that most dense regime systems, with temperatures at the onset of the threshold region ($k_BT \sim E_{\beta}$), are very likely to show collision complex lifetimes that agree with RRKM. However, the ${\rm RbCs}+{\rm RbCs}$ experiment, which is well within this regime, has measured lifetimes that are several standard deviations away in our model. Thus, the RRKM theory, even in this limit, is at best a partial success. 

On the other hand, in the limit of sparse resonances, we have shown that the RRKM result essentially has no meaning in this regime, since there are no nearby resonances over which to average. Instead, we find that time delays are governed by threshold behavior (namely, scattering length) and the strength of the bound-free coupling, with characteristic time scales given by Eq.~(\ref{eq:modified_sparse_Q}).  However, the time delay at threshold includes effects from the long-range potential, making it difficult to associate the delay directly with the collision-complex lifetime. Even when ignoring this subtlety, our model would require coupling strengths that we believe are relatively unlikely in order to match the observed lifetime in the ${\rm KRb}+{\rm Rb}$ experiment. \response{Nonetheless, we do suggest that the scattering length of these molecular species be experimentally measured, in order to confirm whether long-range threshold effects may contribute to long lifetimes.}

\response{Central to our random matrix theory approach is the assumption that each realization of the scattering matrix represents a possible outcome of a close-coupling calculation. If the resonance spectra produced by such calculations obey RMT statistics, then the large mean level spacing in sparse regime systems makes resonances within the experimentally relevant energy window—and hence anomalously long lifetimes—exceedingly unlikely. This suggests that an important source of resonant structure may be absent from our model, which is necessarily limited because we have ignored the details of any particular intermolecular Hamiltonian. Effects such as nonadiabatic couplings, including those associated with conical intersections, quasibound states supported at long range \cite{croft26_longrange}, or quantum scars \cite{kaplan_scars,private} could introduce additional resonances or otherwise enhance the lifetime. To that end, explicitly tracking the multidimensional wavepacket through the collision may help identify how such mechanisms influence the dynamics. Furthermore, recent work has also suggested that a two-body scattering description may be insufficient and that correlated many-body dynamics may need to be considered to address complex lifetimes \cite{peng_zhang}.}


\response{A separate possibility is that the statistical assumptions of the model are incomplete. The actual resonance energies $E_{\mu}$ and bound-continuum couplings $W_{\mu}$ may follow distributions different from those assumed here and may therefore depart from RMT predictions. In addition, we assign the same average coupling strength $x$ to all resonances. If the resonances instead separate into groups associated with different nuclear-spin configurations, each group could be characterized by a different coupling strength and, consequently, a different time scale. How such a hierarchy of couplings affects the connection between stationary resonant scattering and wave-packet evolution remains an open question.}


\response{Despite these limitations, the present work establishes a general distinction between the dense and sparse resonance regimes. In particular, we point out that the RRKM lifetime should not always be used as the reference scale for experimental complex lifetimes, because its underlying resonance-averaging assumption fails in the sparse regime. This conclusion highlights the need for theoretical approaches that incorporate physical mechanisms beyond those included in the present statistical model.}



\section{Data Availability}

The numerical data and code that support the findings of this article are
available from the corresponding author upon reasonable request.

\section{Acknowledgments}
We acknowledge helpful discussions with the K.-K. Ni group and with E. J. Heller, and in the early stages, with J. F. E. Croft and G. Qu\'{e}m\'{e}ner.  This work was supported by NSF grant PHY-2408636, and by the JILA Physics Frontier Center, PHY-2317149.
%

\appendix
\section{Varying the Coupling Strength}
\label{ap:coupling}
\begin{figure}
    \centering
    \includegraphics[width=\linewidth]{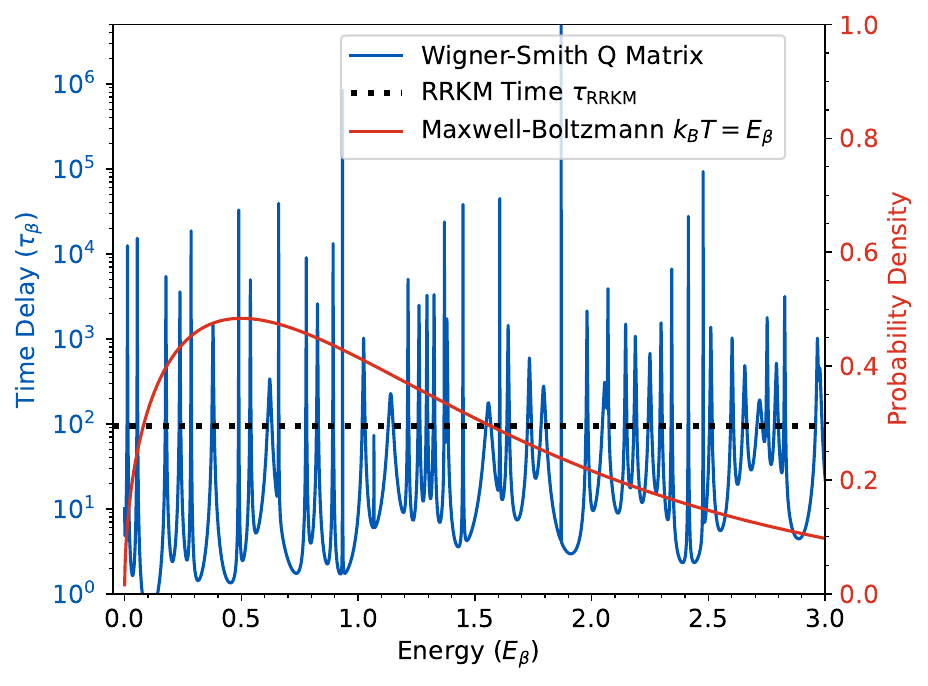} \\
    (a)\\[5pt]
    \includegraphics[width=\linewidth]{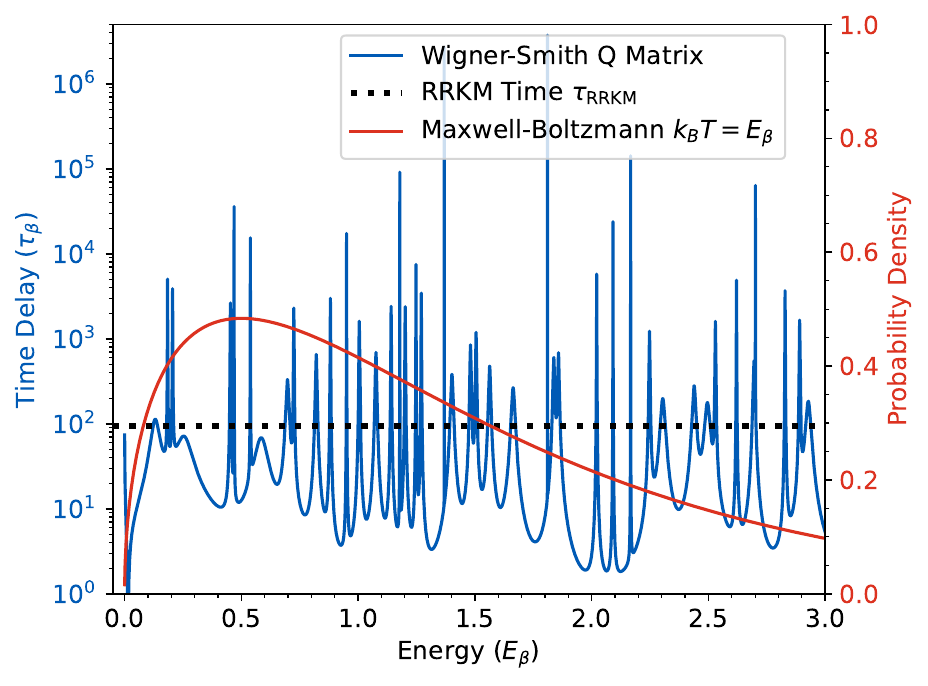} \\
    (b)
    \caption{The time delay $Q(E)$ for (a) $x=0.1$ and (b) $x = 10$. }
    \label{fig:x_dense}
\end{figure}

We examine the effects of varying the bound-free coupling parameter $x$. We plot in Fig.~\ref{fig:x_dense}(a) and (b) the time delay $Q$ for an example spectrum when $x = 0.1$ and $x = 10$ respectively. The predominant effect of shifting $x$ from $1$ is that resonances become sharper. For both $x=0.1$ and $x=10$, the resonances are much narrower on average, and this matches the fact that the theory should be invariant under the transformation $x \mapsto 1/x$ \cite{Mitchell_2010}. In the case when $x=0.1$, the fact that resonances become narrower is not surprising given that the width of resonances scales in proportion to $x$. In the case of $x=10$, overlapping resonances interfere to create a broad envelope and narrow resonances. As a result, the point of largest coupling $x = 1$ is also the point where the resonances in $Q$ can be effectively the broadest. Notably, as the resonances become narrower, they also become taller, such that the integrated contribution from each resonance does not change. Thus, changing the value of $x$ in the dense limit does not affect our analysis in Sec.~\ref{sec:dense}.

\section{Wavepacket Evolution}
\label{ap:wavepacket}
Given that the van der Waals interaction between two molecules is negligible in comparison to the kinetic energy of the wavepacket until fairly short range, the evolution of wavepackets can be approximated as free propagation at long range. This assumption is valid because the phase shift acquired by the actual long-range potential will contribute the same time delay independent of the resonant spectrum.

In order to simulate wavepacket propagation, we suppose that the incoming molecules are described by a Gaussian wavepacket starting at $r_0$ with position uncertainty $\Delta r$ and traveling with center momentum $k_0$ and momentum uncertainty $\Delta k$. The total wavefunction $\psi_{\rm tot} = \psi_{\rm inc} + \psi_{\rm out}$ is given by the sum of the incoming and outgoing wavefunctions
\begin{equation}
\label{eq:wavepacket_prop}
\begin{split}
    \psi_{\text{inc}}(r, t) &= \int \frac{dk}{\sqrt{2\pi}}\ G(k) e^{-i(k^2t + kr)} \\
    \psi_{\text{out}}(r,t, S) &= \int \frac{dk}{\sqrt{2\pi}}\ S(k^2)G(k) e^{-i(k^2t - kr)}
\end{split}
\end{equation}
where $S$ is the scattering matrix for the event and $G$ is the Gaussian envelope
\begin{equation}
    G(k) = \frac{e^{ir_0(k-k_0)}}{\sqrt{\Delta k \sqrt{2\pi}}}\exp\left(-\left(\frac{k-k_0}{2\Delta k}\right)^2\right).
\end{equation}

For convenience, we take the energy corresponding to the center momentum $k_0$ to be the temperature, that is, $E_0 = k_0^2 = k_B T$. It is also natural to consider the width of the distribution of energies to be roughly on the order of $k_B T$ such that $\Delta k \sim \sqrt{k_B T}$. If we take exactly $\Delta k = \sqrt{k_B T}$, however, there is a significant portion of the wavepacket near zero energy, which causes problems in numerical propagation. As a result, we take $\Delta k = \sqrt{k_B T} / 3$ such that zero energy is three standard deviations away. As defined, these wavepackets saturate the uncertainty bound, so this choice of $\Delta k$ also sets the initial positional uncertainty.

\response{The wavepackets are propagated on a positional grid from $-1800\ \beta$ to $2200\ \beta$ using $4000$ grid points. The wavepacket is calculated using a momentum grid of 10000 points with size $8\Delta k$ centered at $k_0$. Time propagation occurs using time steps of roughly $\tau_\beta/(k_0\beta)$ until the wavepacket has exited the collision. Because Eq.~(\ref{eq:wavepacket_prop}) is evaluated directly at each requested time rather than advanced recursively, the sampled time points are independent and can be computed in parallel. Because of the finite momentum grid, resonances which are smaller than the grid spacing cannot be resolved properly. In the case of Fig.~\ref{fig:dense_spectrum}, the energy resolution necessary to resolve the narrowest resonances is roughly $10^{-5}\ E_\beta$, whereas the energy resolution of the wavepacket propagation is only around $5 \times 10^{-4}\ E_\beta$. } 

\section{Measuring Time Delay for Wavepackets}
\label{ap:wp_time_delay}
One way that we can calculate the time delay of wavepackets is by placing a boundary at some radius $r_b$ and measuring the amount of probability density $|\psi|^2$ that has crossed the boundary at time $t$. This can then be integrated to get some characteristic time for when the wavepacket crossed $r_b$. However, there are a few concerns with this approach. One issue is that wavefunction evolution is fairly expensive. The other issue is that when $\Delta k$ is on the order of $k_0$, there is a portion of the wavepacket at arbitrarily small momenta. Combined with the fact that the wavepacket also spreads as it evolves, it can take extremely long times for the entire wavepacket to cross the boundary, at which point numerical errors from the propagation begin to matter. 

Instead, we make use of a mathematical trick to calculate the time delay. Due to the phase factor in $G(k)$, the outgoing wavepacket at $t = 0$ is essentially a wavepacket which is behind the origin $r=0$. This is obviously unphysical since $r$ is technically only defined in the region $[0,\infty)$. However, at large $t$, $\psi_{\text{inc}}$ has now effectively crossed behind the origin to $r < 0$, and $\psi_{\text{out}}$ has traveled into the physical region $r \ge 0$. The physical wavefunction $\psi_{\text{tot}}$ on $r \ge 0$ is then determined mostly by $\psi_{\text{out}}$ after the collision. Thus, we consider the value
\begin{equation}
    \langle r(t, S) \rangle = \int_{-\infty}^\infty dr\ r|\psi_{\text{out}}(r, t, S)|^2
\end{equation}
which gives the center of the outgoing wavepacket for a defined $S$ matrix \textit{even when it is still in the unphysical region}. However, we know that a Gaussian wavepacket travels with group velocity $2k_0$ regardless of the phases, so the difference in $\langle r(t) \rangle$ between the case with resonances and without resonances remains constant to leading order in $t$, which means we can use the difference at $t=0$ to approximate the difference at a later time when $\psi_{\text{out}}$ has entered the physical region. If we then take the difference in $\langle r(t=0)\rangle$ and divide by $2k_0$, this gives a time which corresponds to the delay between a wavepacket with and without resonances. Thus, we define the time delay of the wavepacket to be
\begin{equation}
    \tau = \frac{\langle r(t=0; S=1)\rangle - \langle r(t=0;S=S_Q)\rangle}{2k_0}
\end{equation}
Our numerics furthermore confirm that this approach gives very similar quantities to the procedure described above using the boundary at $r_b$.

\section{Distribution of $\Delta_0$}
\label{ap:wigner}
Let us consider the distribution for $\Delta_0 \equiv E_0-E$ for a fixed energy $E$, where $E_0$ is the nearest energy to $E$ when drawn from the Wigner-Dyson distribution. To do this, we assume a set of energies $\{E_\mu\}$, whose mean level spacing is set by $d$. We denote $E_-$ to be the largest energy from $\{E_\mu\}$ such that $E_- < E$. Then we denote $E_+$ as the smallest energy from $\{E_\mu\}$ such that $E_+ > E$. $E_0$ must therefore be either $E_-$ or $E_+$ by our definition. Given that we have fixed $E$ arbitrarily, we might expect that $(E-E_-)/(E_+-E_-)$ is essentially distributed uniformly between $[0,1]$, that is, $E$ lies randomly in the space between $E_-$ and $E_+$. In that case, we would expect that $E_0 = E + r(E_+-E_-)$ where $r$ is a random variable drawn from the uniform distribution $U(-0.5,0.5)$. To know the statistics of $\Delta_0 = r(E_+-E_-)$, we then only need to know how $E_+-E_-$ is distributed. One might expect that this level spacing should just follow the Wigner-Dyson distribution, but because of the way we picked $E_+$ and $E_-$, it actually follows the distribution given by
\begin{equation}
    P\left(s=\frac{E_+-E_-}{d}\right) = \frac{\pi}{2}s^2 e^{-\pi s^2/4}
\end{equation}
which has $s^2$ scaling as $s\to0$ rather than $s$ scaling as in Eq.~(\ref{eq:wigner-dyson}). Calculating the moments of $\Delta_0 = rsd$ can then be easily done because this is a product of independent random variables $r$ and $s$. Clearly the mean of this distribution is $0$ (as are all odd moments). The variance is given by
\begin{equation}
\begin{split}
    {\rm Var}[rs] &= E[(rs)^2]\\
    &=E[r^2]E[s^2]\\
    &= \frac{1}{2\pi} \approx 0.16
\end{split}
\end{equation}
The fourth moment is 
\begin{equation}
    E[(rs)^4] = E[r^4]E[s^4] = \frac{3}{4\pi^2} = 3({\rm Var}[rs])^2
\end{equation}
As a result, we see that this distribution has the same kurtosis as a normal distribution. Thus, even though the higher standard moments of the distribution may not agree with the normal distribution, we approximate this distribution as a normal distribution with mean $0$ and variance $0.16d^2$ so that we can use analytic formulas in our derivations.\\[5pt]

\response{
\section{Distribution in the Sparse Limit}
\label{ap:sparse}
We want the distribution of $\alpha_0 = W_0^2/\Delta_0$, where $W_0$ and $\Delta_0$ are drawn from Eq.~(\ref{eq:distribution}). First, we define $\sigma_\Delta^2 = 0.16 d^2$ and $\sigma_W^2 = xd/\pi^2$. The distribution for $y=\Delta_0$ is therefore
\begin{equation}
    P_y(y) = \frac{1}{\sqrt{2\pi \sigma_\Delta^2}} \exp\left[-\frac{y^2}{2\sigma_\Delta^2}\right].
\end{equation}
Now, we consider that since $W_0$ is normally distributed with variance $\sigma_W^2$, then the distribution of $W_0^2$ is the square of a Gaussian distribution. It turns out that the square of the normal distribution with variance $1$ is given by the chi-squared distribution $\chi_1^2$. As a result, the distribution for $w = W_0^2$ is given by
\begin{equation}
    P_w(w) = \frac{1}{\sqrt{2\pi w\sigma_W^2}} \exp\left[-\frac{w}{2\sigma_W^2}\right]
\end{equation}
for $w \ge 0$. To find the distribution of $\alpha_0$, we must then take the ratio distribution of $z = w/y$. Since these variables are independent, this can be done by taking the integral 
\begin{equation}
\begin{split}
    P_z(z) &= \int_{0}^\infty dy\ |y| P_y(y) P_w(zy)\\
    &= \frac{1}{2\pi \sigma_W \sigma_\Delta \sqrt{z}} \int_{0}^\infty dy\ \sqrt{y} \exp\left[-\frac{y^2}{2\sigma_\Delta^2} - \frac{zy}{2\sigma_W^2}\right],
\end{split}
\raisetag{50pt}
\end{equation}
where we have focused only on the positive branch of $z \ge 0$, noting that $P(z) = P(-z)$. Since $W_0^2 = zy$, it must be that $y \ge 0$, which is why the bounds of integration run from $0$ to $\infty$. Now, substituting the variances, letting $c=0.4$, and making the replacement $u = y / d$ yields
\begin{equation}
    P_z(z) = \frac{1}{2c\sqrt{xz}} \int_0^\infty du\ \sqrt{u}\exp\left[-\frac{u^2}{2c^2} - \frac{\pi^2zu}{2x}\right].
\end{equation}
It can therefore be immediately seen that this distribution does not depend on the value of $d$. Evaluating this integral produces the final result
\begin{equation}
\begin{split}
    P_z(z) &= \frac{\lambda |z|}{\pi}e^{\lambda z^2}\left[K_{3/4}(\lambda z^2) - K_{1/4}(\lambda z^2)\right] \\
    \lambda &= \frac{\pi^4c^2}{16x^2}.
\end{split}
\end{equation}
where $K_{\alpha}$ denotes the modified Bessel function of the second kind of order $\alpha$. Although $P_z$ is symmetric around 0, the resulting probability density function is both heavy-tailed and diverges at $0$, so there is no defined standard deviation or full width at half maximum. We thus characterize the width of the distribution by taking the value $\gamma$ at which exactly $50\%$ of the distribution falls between $-\gamma$ and $\gamma$. Numerically calculating this gives $\gamma = 0.18x$.}

\end{document}